\documentclass[prl,twocolumn,10pt]{revtex4}

\usepackage{graphicx}
\usepackage{dcolumn}
\usepackage{bm}
\usepackage{times}

\begin{document}


\title{
Electronic features around Fermi level correlating to occurrence
of magnetism or superconductivity in Laves-phase intermetallic
compounds}

\author{Xing-Qiu Chen$^{a}$}
\author{Shu-Lan Wang$^{b}$}
\author{Xue-Yong Ding$^{a}$}
\author{Xiang-Xun Xue$^{a}$}

\affiliation{%
$^a$ School of Materials and Metallurgy, Northeastern University,
Shenyang 110004, Peoples' Republic of China,}
\affiliation{%
$^b$ School of Science, Northeastern University, Shenyang, 10004,
Peoples' Republic of China. }

\begin{abstract}
Based on density functional calculations, the relationship between
magnetism or superconductivity and electronic states around Fermi
level were derived, and the location of the Fermi level in
nonmagnetic "form" of Laves-phase compounds is very sensitive to
determine the presence of nonmagnetism, or ferromagnetism, or
antiferromagnetism, or superconductivity. The Fermi level at the
nonbonding and antibonding regions corresponds for the
nonmagnetism and magnetism, respectively, whereas at bonding
region with responsibility for superconductivity. This rules will
be very useful and convenient to search the new magnetic and
superconductive materials.
\end{abstract}
\keywords{Electronic states at Fermi level; magnetism,
superconductivity}
\maketitle

The searches of new magnetic and superconductive materials have
always simulated extensively interesting in materials science. As
we know, a precise location of $E_F$ is very important for those
physical properties (such as magnetism and superconductivity)
which depend critically on the value of the DOS at Fermi level.
Generally, in the nonmagnetic phase the higher values the DOS at
Fermi level have, the more possible the compound is magnetic or
superconductivity. Nowadays, for magnetic systems most solid-state
physicists accept the Stoner criterion \cite{Stoner} to decide
whether or not a metal or alloy should be ferromagnetic,
\begin{equation}
IN(E_F) > 1
\label{Estoner}
\end{equation}
where $I$ is a measure of the strength of the exchange interaction
in the metal and $N(E_F)$ is the density of states at Fermi level,
$E_F$. As a predictive tool Stoner theory is actually not always
successful, for instance, in the first transition metals Fe, Co,
Ni are all ferromagnetic, however using Stoner theory for cobalt
$IN(E_F)$ = 0.972 $<$ 1 \cite{Janak} which indicated Co should be
nonmagnetic; It is even also somewhat strange that nickel, the
weakest ferromagnetic, has the largest value $IN(E_F) \approx$ 2.0
\cite{LD}.
On the other hand, for
superconductive materials a basic ingredient in theories
evaluating the superconduction transition temperature is the
electron-phonon interaction \cite{Huang}. The strength of this
interaction is conveniently expressed by $\lambda$, the
electron-phonon coupling parameter,
\begin{equation}
\lambda = \frac{N(E_F)<I^2>}{M<\omega^2>}
\label{stoner}
\end{equation}
where $<I^2>$ is the electronic stiffness parameter of Gaspari and
Gyorffy \cite{Gaspari}, $M$ the ionic mass, and $<\omega^2>$ some
mean value of the phonon frequency. From the above equations we
have deliberately used definitions from the simplest available
electronic theories for magnetism and superconductivity, just to
stress the relevant role played in these phenomena by $N(E_F)$.
For example, HfV$_2$, with high DOS of V atom
satisfying the Eq.\ref{Estoner},
is nonmagnetic superconductive materials.
Then, at least high
DOS at Fermi level does not guarantee the occurrence of
ferromagnetism or superconductivity.
Therefore, it is very necessary
to pursuit the more reliable tool to estimate whether or not a
material is ferromagnetic, or antiferromagnetic, or
superconductivity.
For this point,
in 2000 Landrum and Dronskowski \cite{LD} made
the first work --
they thought whether or not an alloy contains ferromagnetic
elements, the presence of antibonding states at $E_F$ serves
as a "fingerprint" to indicate a ferromagentic instability.

In this Letter, we present an understanding for occurrence of
magnetism or superconductivity in Laves-phases compounds by the
electronic states around Fermi level. Laves phase compounds, the
largest class of intermetallics, have a wide interesting physical
and technological properties in particular for magnetism,
superconductivity, and hydrogen storage
\cite{zrzn2,zrzn23,nat1,nat2}. Many of these compounds crystallize
in the densely packed hexagonal C14 and C36, and the cubic C15
crystal structures \cite{laves}. Searching for the ground state
structure of Mn-, Fe-, Cr-, V-based binary transition-metal Laves
phases (total 16 compounds) we performed density functional
calculations for the C14, C15 and C36 structures finding many new
and interesting magnetic ground and metastable states
\cite{Xingqiu}: for Mn-based Laves phases, ScMn$_2$, YMn$_2$, and
LuMn$_2$ crystallize in C15 antiferromagnetic (AF) ground states,
and TiMn$_2$ \cite{Chen3}, HfMn$_2$ and NbMn$_2$ have C15
ferromagnetic (FM) metastable states, and ZrMn$_2$ stabilizes the
C15 FM with near-degeneracy of C15, C14, and C36 phases at ground
state \cite{Chen2}. For TiFe$_2$, ZrFe$_2$, and HfFe$_2$, all of
them crystallize C15 FM ground states; RCr$_2$ and RV$_2$ (R=Ti,
Zr, Hf) have all C15 nonmagnetic (NM) ground states. When we
analyzed the electron states of these C15 stable or metastable
Laves phases, a principle can be found for NM, AF, FM, and
superconductivity. This new finding is the subject of our paper.

For the calculation of ground state properties of
these above mentioned Laves phase compounds in the
C14, C15 and C36 crystal structures we applied density functional
theory (DFT) by means of the plane wave Vienna Ab initio
Simulation Package (VASP)\cite{vasp} which -in its projector augmented wave
formulation
- is one of the most precise methods for calculating the energetics and
 electronic structure of solid matter within periodic boundary conditions.
For the many-body exchange-correlation interaction the generalized
gradient approximation of Perdew and Wang \cite{gga91} in
combination with the approach of Ref. \cite{vosko} for spin
polarization was chosen.  The calculations are free from any
empirical parameters. Lattice parameters as well as the atomic
coordinates are determined by minimizing total energies and atomic
forces. Care was taken that for each structure a sufficient number
of {\bf k} points for the Brillouin zone integration was chosen.
Ferromagnetic as well as some selected antiferromagnetic spin
arrangements were considered.

\begin{figure}[htbp]
\begin{center}
 \includegraphics[width=2.0in,angle=0]{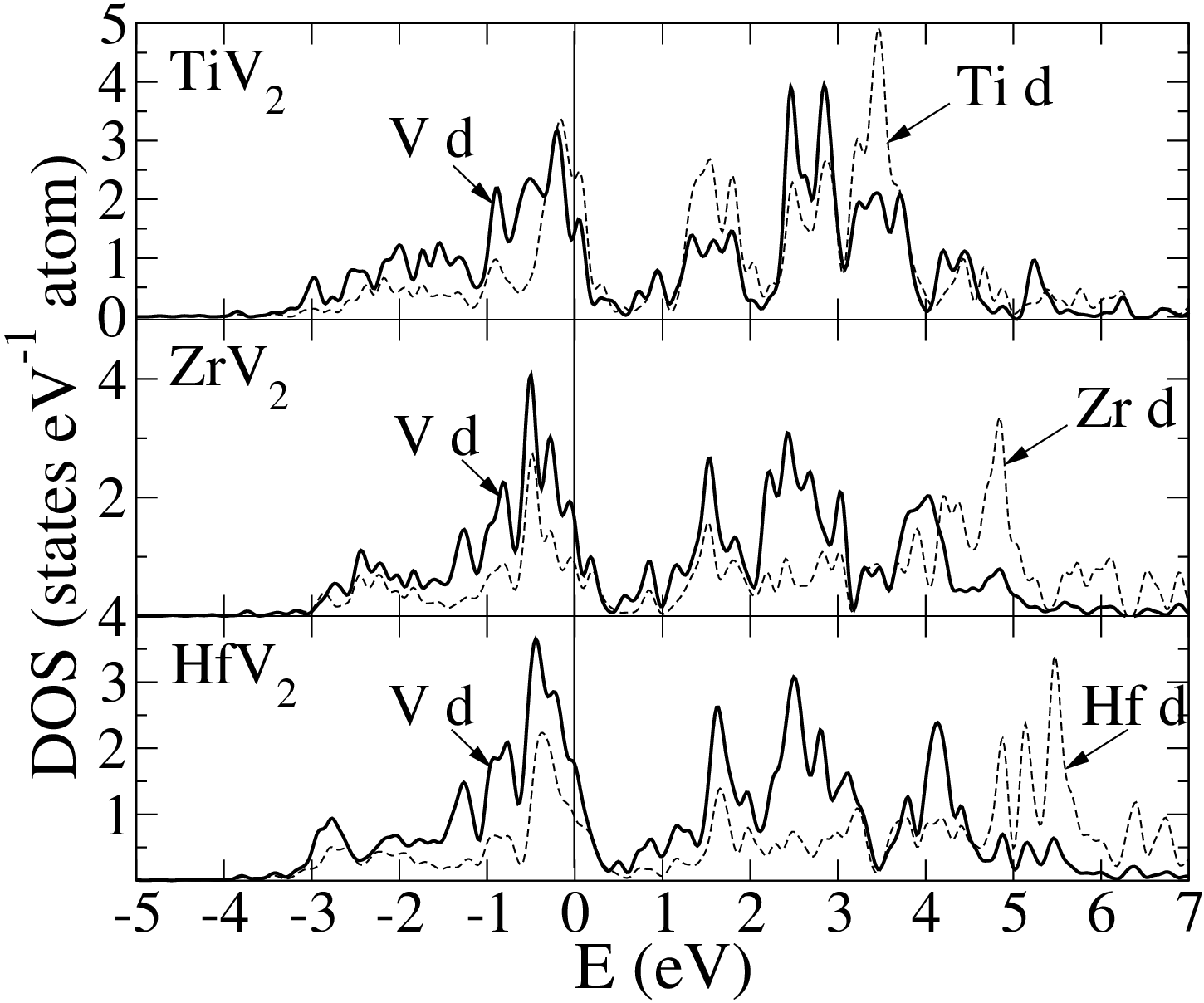}\\
 \includegraphics[width=2.0in,angle=0]{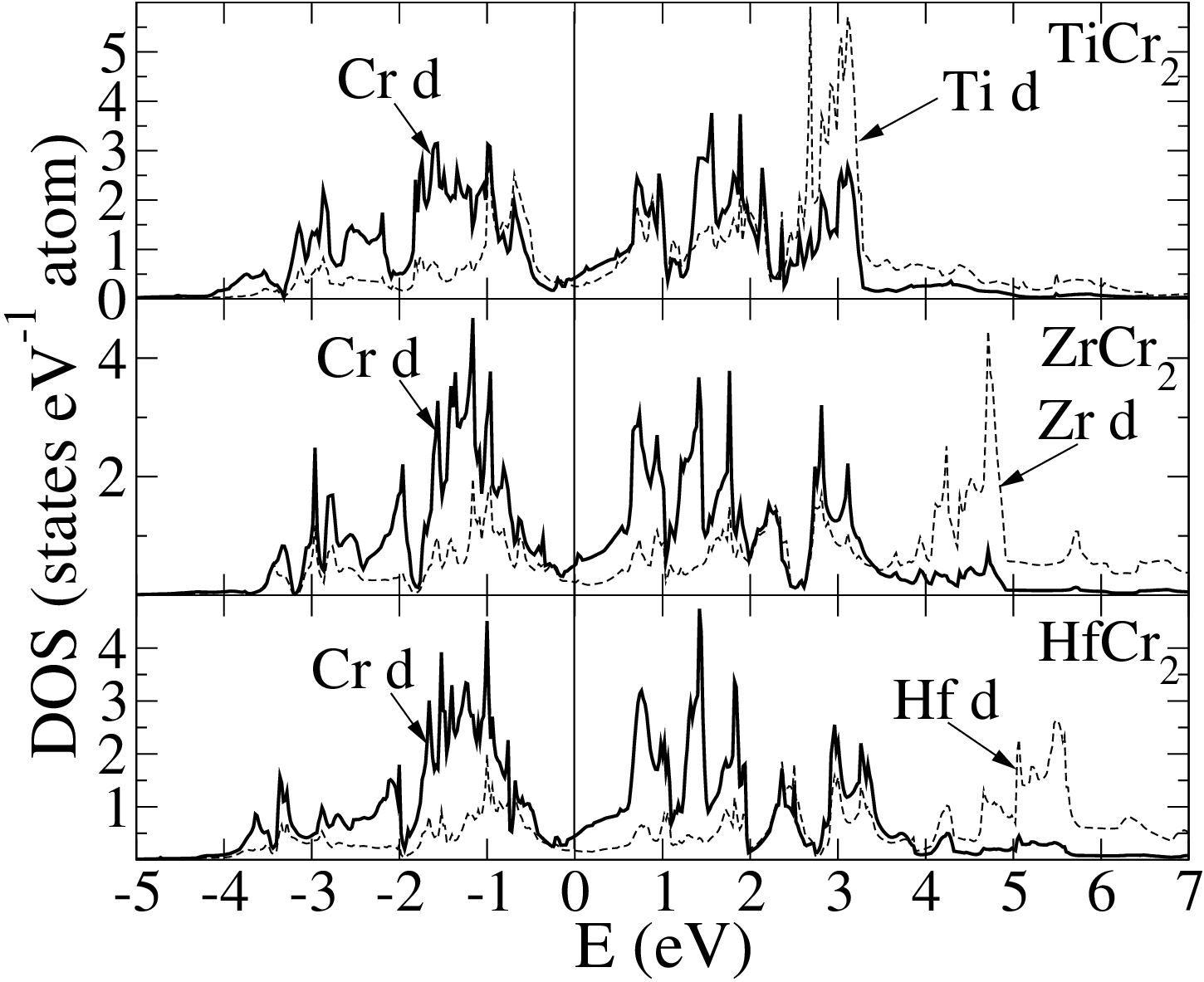}\\
 \includegraphics[width=2.0in,angle=0]{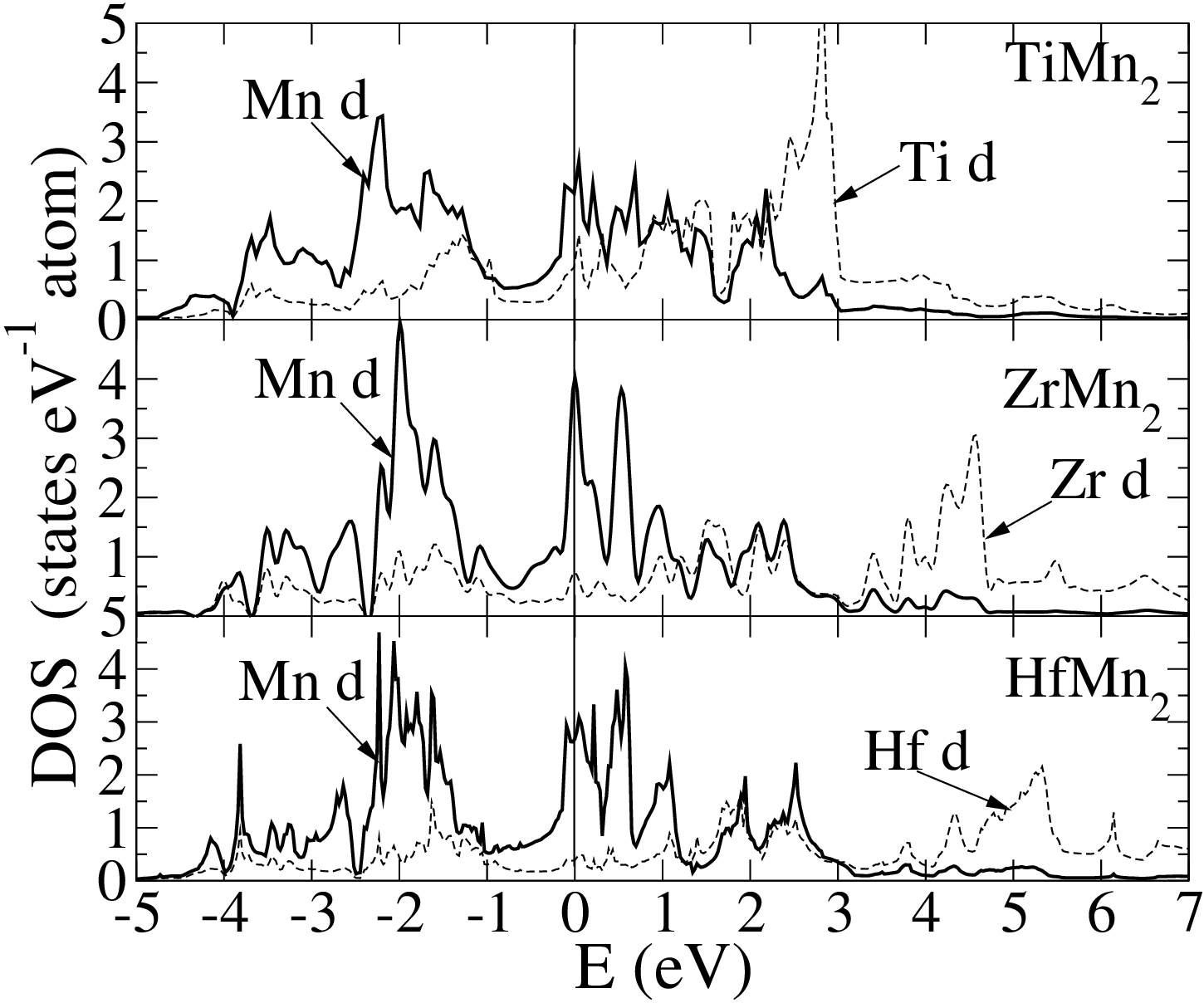}\\
 \includegraphics[width=2.0in,angle=0]{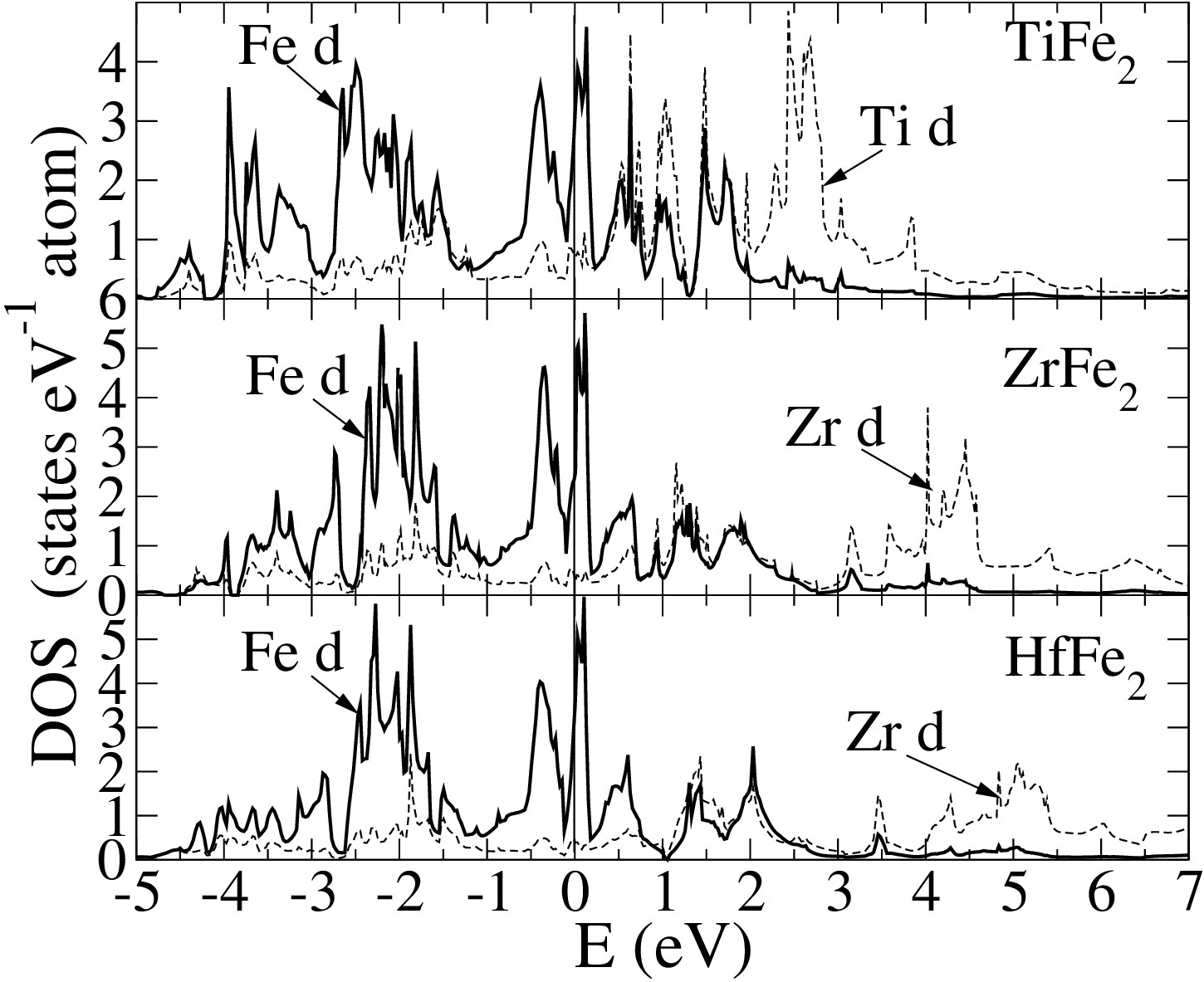}\\
\caption{ Density of states (DOS) of local $d$-like of C15
nonmagnetic RCr$_2$, RV$_2$, and RMn$_2$, and RFe$_2$ (R=Ti, Zr,
Hf). Thick solid curves are corresponding for Cr, V, Mn, and Fe
atoms; the dotted curves are the DOS of R atoms. \label{fig1}}
\end{center}
\end{figure}

The electronic structures at their equilibrium volumes of
nonmagnetic Laves phase compounds are shown in Fig.\ref{fig1}. The
total DOS curves (not shown here), which runs from -5 to 7 eV, are
mainly composed of $d$ states of both two constituent elements in
compound RM$_2$ (R = Ti, Zr, Hf; M = V, Cr, Fe, Mn). The distinct
features for all 16 compounds is that the pronounced pseudogap
around Fermi energy occurs separating the energy regions of
bonding and antibonding states. A very large peak of unoccupied
d-like states of R atoms arises at high energy part above Fermi
level which is due to nonbonding states. In compounds RV$_2$
(R=Ti,Zr,Hf) the Fermi level $E_F$ falls in a relative high DOS at
down-shoulder of strong R-V {\em bonding} region, whereas for
RCr$_2$ the Fermi level lies the valley of the pronounced
pseudogap with very low DOS values. Clearly, RV$_2$ compounds have
relative high DOS at Fermi level, even satisfying Stoner theory.
So, one would expect the occurrence of ferromagnetic phases or the
instability of structure. However, our spin-polarized calculations
revealed that they are nonmagnetic. Furthermore, experiments found
that HfV$_2$ and ZrV$_2$ are superconductive materials with C15
structures (ZrV$_2$ with $T_{sc}$ = 8.8 K \cite{Matthias}; HfV$_2$
with $T_{sc}$ = 9 K \cite{Luthi}). Up to date, no experiment
reported C15-type TiV$_2$. The calculated enthalpies of formations
of 30.2, -3.6, and -4.5 kJ/mol for TiV$_2$, ZrV$_2$, and HfV$_2$,
respectively, prove that C15 TiV$_2$ is unlikely to exist
thermodynamically due to highly positive formation energy. On the
contrary, according to the theory of electronic structure the low
DOS value at Fermi level means the stability of system. Therefore,
in RCr$_2$ compounds the appearance of $E_F$ in low pseudogap
indicates a system is stable with C15 NM phase, which are perfect
agreement with experimental investigations \cite{Massalski}.

\begin{figure}[htbp]
\begin{center}
 \includegraphics[width=2.0in,angle=0]{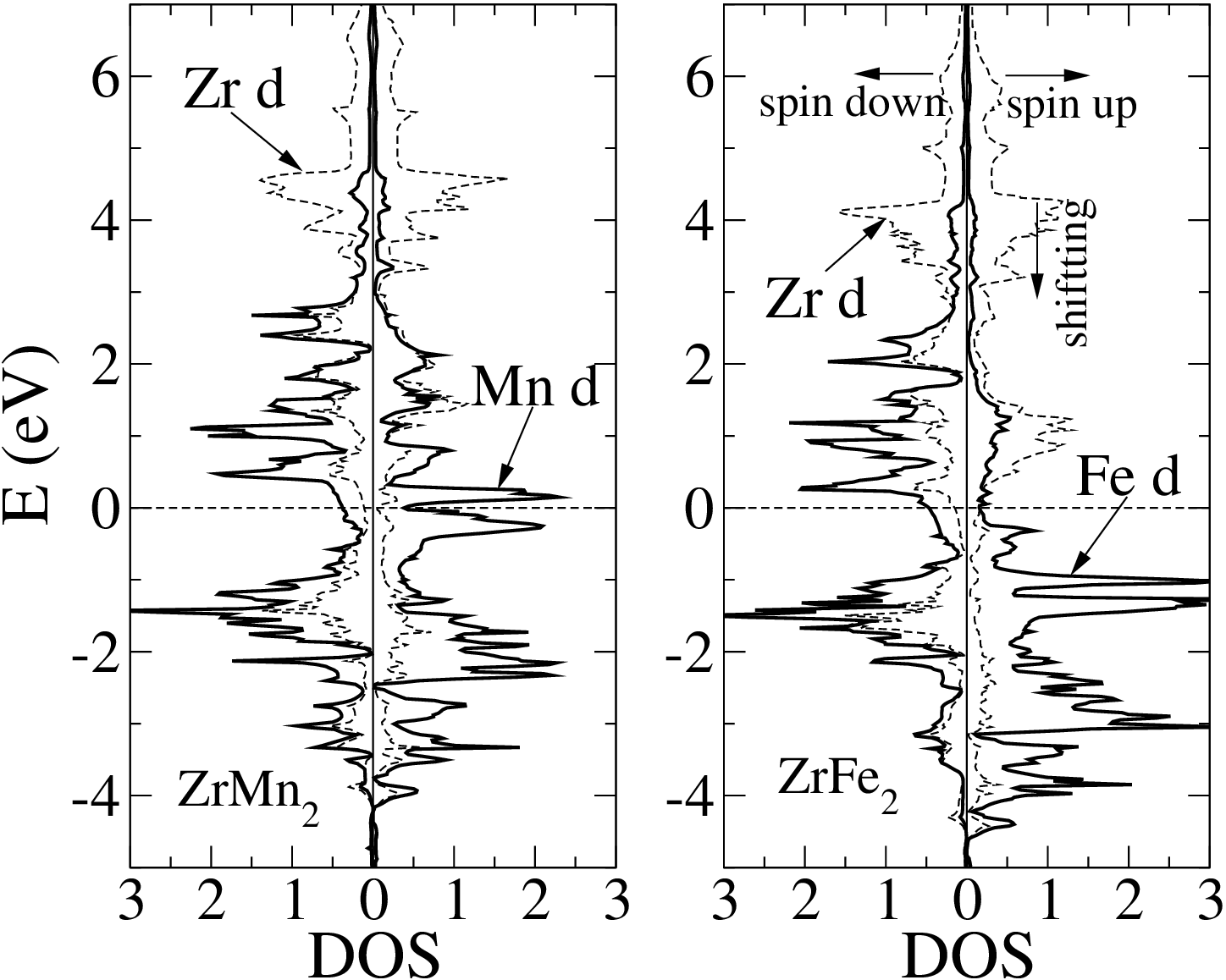}\\
\caption{
Spin-polarized magnetic spin up and spin down
densities of states of $d$-like states in C15 FM ZrMn$_2$
and ZrFe$_2$. The solid thick black curves are the DOSs of
d-like state of Mn and Fe atoms; the dotted curves are the
DOSs of Zr atoms in both cases.
\label{fig2}}
\end{center}
\end{figure}

For RMn$_2$ and RFe$_2$ compounds, DOS curves are similar to
RCr$_2$ and RV$_2$. The important difference is that the Fermi
level now lies on the sharp peak of {\em antibonding} region. It
is also noted that there is an obvious nonbonding region between
R-$d$ and Mn- (Fe-)$d$ states below the Fermi level, where
$d$-like states of all R atom have a very flat curves around the
pseudogaps (see Fig.\ref{fig1}). Actually, our further
spin-polarized calculations show that these nonmagnetic C15
RMn$_2$ and RFe$_2$ compounds are unstable with respect to an
electronic structure distortion. The magnetic calculations
revealed that all these compounds have C15 ferromagnetic ground or
metastable states (their magnetic properties are given in Table
\ref{tab1}, specially in which metastable state means that the
compounds have C14 stable ground state, but their C15 phase have
negative formation energy. One would expect that the phase
transition occurs from C14 to C15 type, for example, ZrMn$_2$
\cite{Chen2} and TiMn$_2$ \cite{Chen3}). Clearly, the
ferromagnetic moments of the compounds RMn$_2$ are much less than
those of RFe$_2$ compounds. This difference can be attributed to
the strong ferromagnetism of pure Fe metal and very weak
antiferromagnetism of pure Mn metal. However, it must be pointed
out that to date in these compounds only ZrFe$_2$ has been found
experimentally to be C15 FM with the local moment 2.0 $\mu_B$/Fe
\cite{zrfe2}, which agrees perfectly with current calculations.
Certainly, with respect to their nonmagnetic "forms", a
rearrangement of electron spins in magnetic systems are caused by
the spontaneous magnetization. Namely, the spin up and spin down
sublattices become inequivalent and shift in energy (see
Fig.\ref{fig2}). As example, we only give the spin DOSs of C15 FM
ZrMn$_2$ and ZrFe$_2$ (other compounds also have similar DOS
features, not shown here). We found that the spin-polarization has
not significantly affected the shape of the DOSs; merely shifted
them in energy. For C15 ZrFe$_2$ the spin-up DOS has decreased in
energy enough to almost completely fill the Fe $d$-like states
(see thick solid curves in Fig.\ref{fig2}). Also in C15 ZrMn$_2$
cases the spin-up DOS of $d$-like states of Mn atoms shift up to
almost make the band fully unoccupied, however the spin-down
electrons states must decrease to make more electrons occupy this
state. This kind of splitting in energy for spin-up and spin-down
states is namely the exchange spin splitting (these values of both
RMn$_2$ and RFe$_2$ are also given Table \ref{tab1}), which result
in the occurrence of ferromagnetism for these compounds. Now we
still see another important key indications that the Fermi level
$E_F$ falls down the valleys (very low DOS, see Fig. \ref{fig2})
of DOS curves in magnetic cases - imply that these compounds can
crystallize in C15 FM phases.

\begin{table}[htbp]
\begin{center}
\caption{The stabilities and magnetic properties
of C15
RMn$_2$ (R=Sc, Y, Lu; Ti, Zr, Hf)
and RFe$_2$ (R=Ti, Zr, Hf) Laves
phase compounds.
\label{tab1}}
\begin{ruledtabular}
\begin{tabular}{ccccc}
Compounds & stabilities & moment ($\mu_B$) & splitting (eV)\\
\hline
TiMn$_2$ & metastable FM  & 0.76/Mn  & 0.78\\
ZrMn$_2$ & ground state FM & 0.90/Mn & 0.94\\
HfMn$_2$ & metastable FM  & 0.86/Mn  & 0.90\\
TiFe$_2$ & ground state FM & 1.5/Fe  & 1.56\\
ZrFe$_2$ & ground state FM & 2.0/Fe  & 2.03\\
HfFe$_2$ & ground state FM & 1.3/Fe  & 1.32\\
ScMn$_2$ & metastable AF   & $\pm$1.4/Mn & 1.42 \\
YMn$_2$ & ground state AF  & $\pm$2.6/Mn & 2.67 \\
LuMn$_2$ & ground state AF & $\pm$2.2/Mn & 2.22 \\
\end{tabular}
\end{ruledtabular}
\end{center}
\end{table}

\begin{figure}[htbp]
\begin{center}
 \includegraphics[width=2.0in,angle=0]{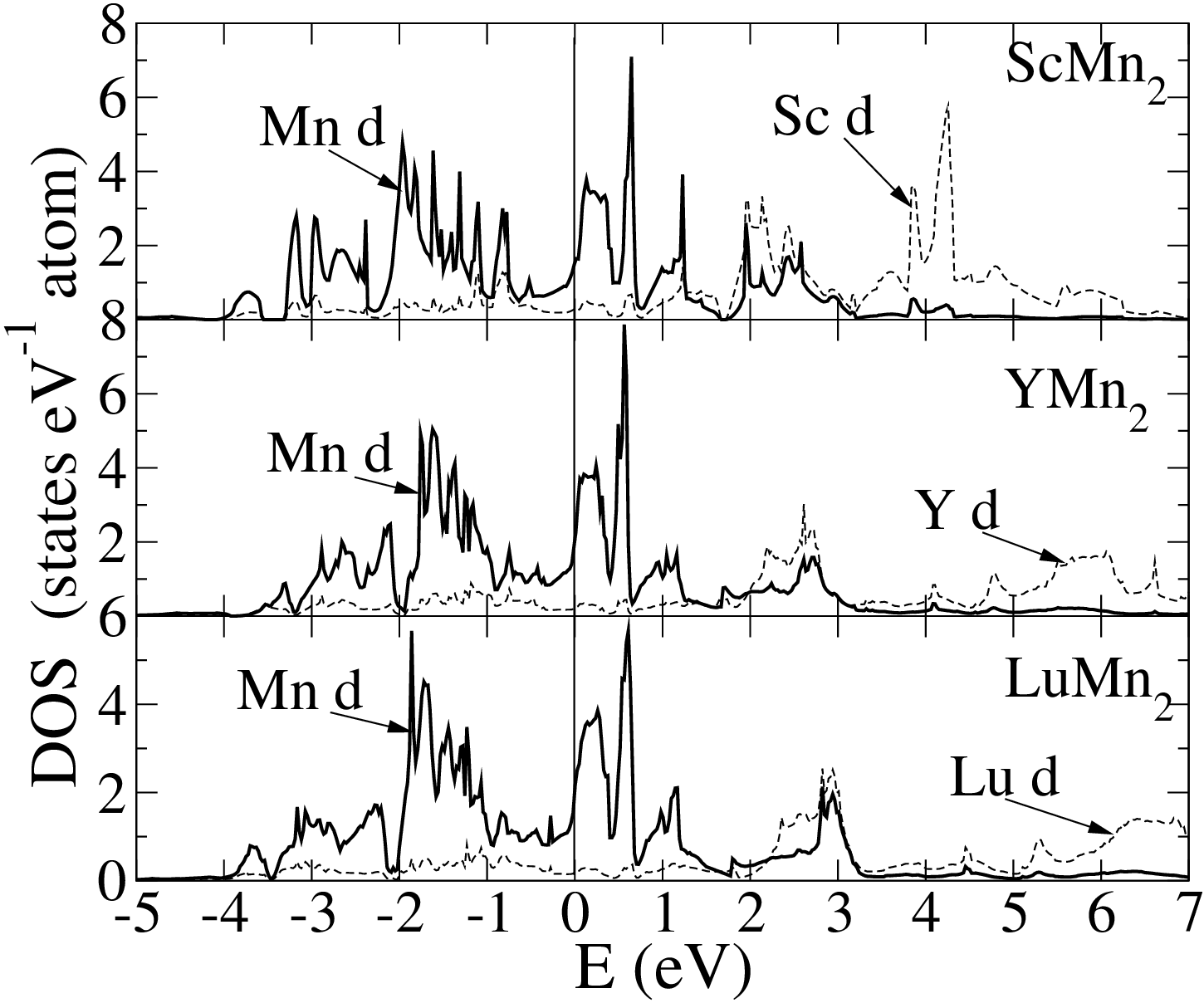}\\
\caption{
Density of states of local $d$-like of C15 nonmagnetic
ScMn$_2$, YMn$_2$, and LuMn$_2$.
\label{fig3}}
\end{center}
\end{figure}

\begin{figure}[htbp]
\begin{center}
\includegraphics[width=2.0in,angle=0]{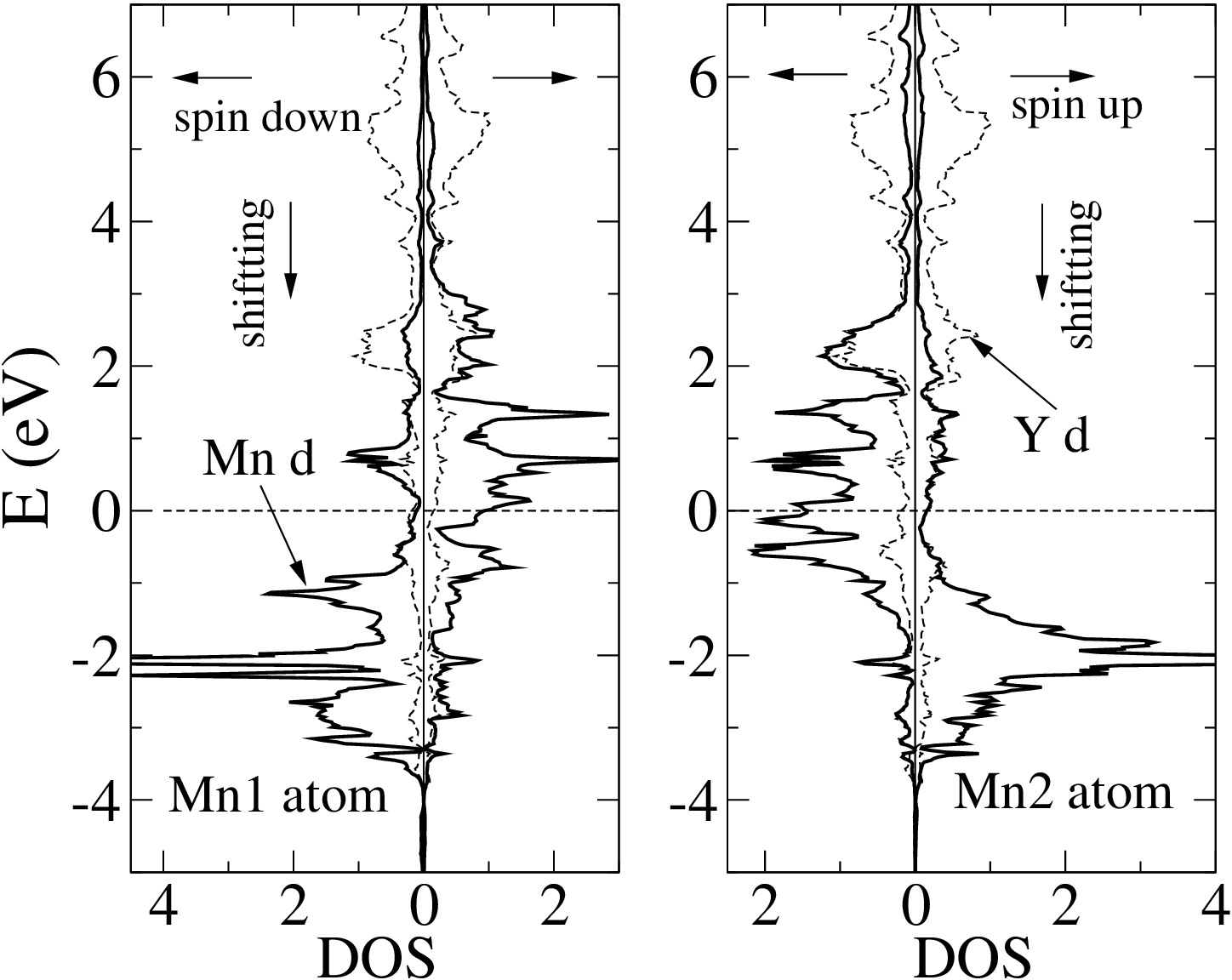}\\
\caption{
Density of states of local $d$-like of C15 AF
YMn$_2$. Mn1 atom is the Mn atom with negative magnetic moment;
and Mn2 atom stands for the Mn atom with
the positive magnetic moment.
\label{fig4}}
\end{center}
\end{figure}

Furthermore, RMn$_2$ (R=Sc, Y, Lu) compounds with C15 Laves phase
type are also investigated. The electronic states of them are
shown in Fig.\ref{fig3}. It is very clear to see that $E_F$ lies
in the {\em crossover point between nonbonding and antibonding
regions} with high DOS values in all cases. At the left of the
Fermi level there are also pronounced pseudogap with very flat
curves of $d$-like states of R atoms, and at the right of the
Fermi level a relative weak antibonding region begin to appear.
Because their DOS values at Fermi level satisfy the Stoner theory,
the C15 phases of these compounds should be expected to have
ferromagnetic properties as well. Surprisingly, our spin-polarized
calculated revealed that their C15 phases are not all
ferromagnetism, but {\em antiferromagnetism} (local moments are
listed in Table \ref{tab1}). By spontaneous AF spin polarization
all three compounds gain different magnetization energies defined
by $U_{DFT}(AF;V_0)-U_{DFT}(NM;V_0)$ are 2.4, 26.1, and 7.7 kJ/mol
with respect to their nonmagnetic states for ScMn$_2$, YMn$_2$,
and LuMn$_2$, respectively. Additionally, the happy fact is that
for YMn$_2$ our current calculations are well agreement with the
recent experimental observations (2.7 $\mu_B$/Mn with C15 AF
structure \cite{ymn2}). For ScMn$_2$ and LuMn$_2$ no comparison of
magnetic structures is possible due to the lack of experimental
measurements. Now we briefly described the antiferromagnetic
electronic states of these compounds. Here, we only gave out the
DOS of C15 AF YMn$_2$ in Fig. \ref{fig4} (ScMn$_2$ and LuMn$_2$
are similar to YMn$_2$, not shown here). Mn atoms in C15 AF unit
cell have two different DOS curves, but only the fully opposite
between Mn1 and Mn2 atoms. Unlike ferromagnetism, the spin
ordering of Mn atoms must be coupled antiparallel to each other --
lead to the total magnetic moment is zero. The occurrence of
magnetism for each Mn atoms is caused by the exchange splitting
(Table \ref{tab1}) in energy (see Fig. \ref{fig4}) upon the
spontaneous spin polarization.

So, Now we revived again to note the electronic states of all
RM$_2$ Laves phase compounds considered here according to above
discussions. There are some important and simply features can be
drawn for
their nonmagnetic electronic states:\\
(i) for nonmagnetic Laves compounds, e.g. RCr$_2$ (R=Ti, Zr, Hf),
the Fermi level $E_F$ lies in the valley of the pronounced
pseudogap.;\\
(ii) for superconductivity Laves compounds, e.g. ZrV$_2$ and
HfV$_2$, the Fermi level $E_F$ lies in the {\em bonding} region
with relative high DOS value;\\
(iii) for ferromagnetic Laves compounds, e.g. ZrFe$_2$, ZrMn$_2$,
etc, the Fermi level $E_F$ lies in the strong {\em antibonding}
region with very high DOS value (generally sitting at a sharp peak);\\
(iv) for antiferromagnetic Laves compounds, e.g., RMn$_2$ (R=Sc,
Y, Lu), the Fermi level $E_F$ is close to the {\em crossover
boundary} between nonbonding and antibonding regions also with
high DOS values.

\begin{figure}[htbp]
\begin{center}
\includegraphics[width=2.0in,angle=0]{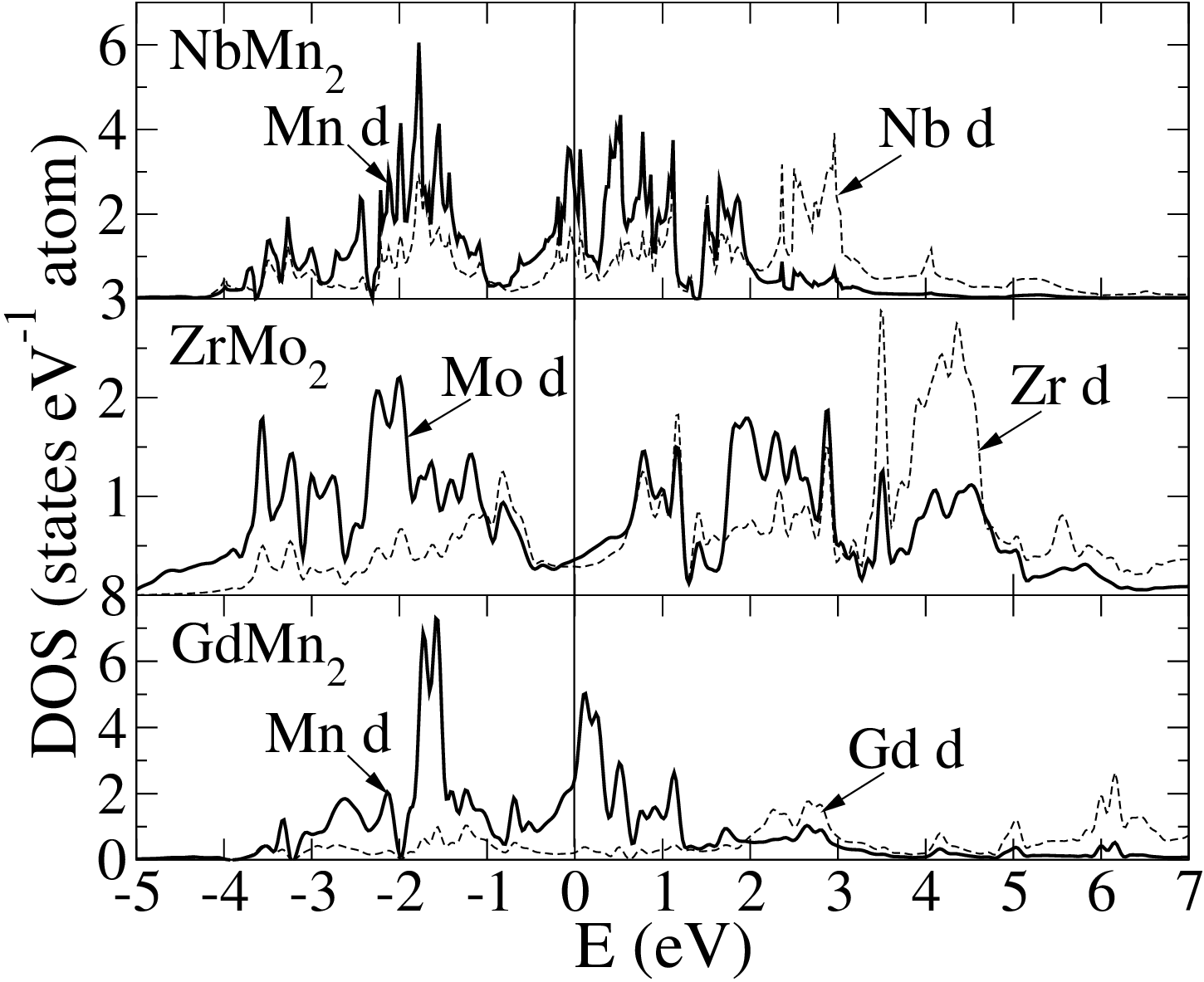}\\
\caption{
Density of states of local $d$-like of C15 NM
NbMn$_2$.
\label{fig5}}
\end{center}
\end{figure}

Currently, for TiV$_2$ the above analysis indicated that it
satisfies the criterion (ii), then it should be superconducting
materials. But, as already pointed out, it is a pity that the
phase actually is unstable due to highly positive enthalpy of
formation. In order to check the reliability of the principles, we
also further investigated ZrMo$_2$, NbMn$_2$, and GdMn$_2$ Laves
phase compounds with C15 phase. It has been found C15 phase of
NbMn$_2$ is metastable and ZrMo$_2$ and GdMn$_2$ have C15 ground
state. The electronic structures in the C15 "nonmagnetic" forms
are given in Fig.\ref{fig5}. Clearly, it can be seen that the
Fermi level $E_F$ of C15 NbMn$_2$ falls on the sharp peak in the
strong {\em antibonding} region between Nb-$d$ and Mn-$d$ states
(satisfying the criterion (iii)). Then the phase should be
ferromagnetism. The spin-polarized calculations exactly reveal the
point -- Mn atom has a moment of 0.6$\mu_B$ in C15 NbMn$_2$; For
C15 ZrMo$_2$, the Fermi level lies on the valley of pseudogap
where the obvious {\em nonbonding} region appears (satisfying the
criterion (i)), meaning it should be the stable nonmagnetic phase.
The spin-polarized calculations definitely found that it is
nonmagnetic; for C15 GdMn$_2$, from Fig. \ref{fig5} a {\em
nonbonding} region of Mn-$d$ and Gd-$d$ states exists from -0.6 eV
to $E_F$, and at the right side of $E_F$ the {\em antibonding}
region begin to appear. Namely, the Fermi level $E_F$ exactly sits
at the {\em crossover boundary} between nonbonding and antibonding
regions (satisfying the criterion (iv)). So the compound GdMn$_2$
should have antiferromagnetic property. The spin-polarized
calculations revealed that it is an antiferromagnetic structure
with local moments of $\pm$2.6 $\mu_B$ per Mn atom and
$\pm$0.4$\mu_B$ per Gd atom. From experiments \cite{Ouladdiaf} it
can be found that the local moments are $\pm$ 2.1 and $\pm$ 4.6
$\mu_B$ for Mn and Gd atoms, respectively. The difference of the
local moment between the calculations and experiments for Gd atom
is mainly due to the soft potential (calculating nine valence
electrons) of Gd atom applied in the calculations. Normally, I
would use the hard potential for Gd (with 18 valence electrons).
Unfortunately, VASP does not converge for the hard potential in
GdMn$_2$ case.

At last, the emphases should be played on the reliability of
predictions in this work. The properties of all these compounds
were predicted based on the atomic scale simulations within
quantum physical density function theory. Many established facts
are the right ground states and phase relation for these compounds
are predicted. On the other hand, the calculations of ZrFe$_2$ and
YMn$_2$, which are in perfect agreement with experimental results,
also further confirmed our results. For all compounds considered
here, the obtained the equilibrium lattice constants and lattice
parameters are in good agreement with experiment
\cite{Xingqiu,Chen3,Chen2}. The calculated volume is slightly
smaller within 5\% due to the approximations made for the
many-body term of DFT. Furthermore, the quality of our data is
corroborated by the calculated energies of formation which in many
cases agree with reliable experimental data, such as TiMn$_2$
\cite{Chen3}.

Summarizing, our argument is that the occurrence of
magnetism and superconductivity in Laves phases compounds directly
depended on the electronic states around Fermi level in their
nonmagnetic phase. The basis principles of (i), (ii), (iii), and
(iv) have already been established to open the
useful clues for searching new
magnetism and superconductive materials.

\end{document}